\documentclass{appolb}
\usepackage{graphicx}

\def\ba{\begin{eqnarray}}
\def\ea{\end{eqnarray}}
\def\lb{\label}

\begin{document}
\title{Blast-wave model description of the HBT radii measured in $pp$ collisions at the LHC energies%
\thanks{Presented at the XI Workshop on Particle Correlations and Femtoscopy, November 3--7, 2015, Warsaw, Poland}%
}
\author{Wojciech Florkowski
\address{Institute of Nuclear Physics, 31-342 Krakow, Poland}
\\
}
\maketitle
\begin{abstract}
Blast-wave model is applied to describe the Hanbury-Brown--Twiss (HBT) radii of pionic systems produced in $pp$ collisions at the LHC energies.
\end{abstract}

\medskip
\centerline{\it Dedicated to Professor Jan Pluta on the occasion of his 70th birthday}

\section{Introduction}  
At very high energies, such as those presently available at the LHC, the final state hadron multiplicities in $pp$ collisions become large and  hydrodynamic description of hadron production in these processes may become possible in the similar way as in peripheral heavy-ion collisions~\cite{Ferroni:2011fh,Bozek:2013ska,Ghosh:2014eqa}. Following this expectation, in this conference proceedings I use the hydro-inspired, blast-wave model to analyse the ALICE data on pion correlation functions~\cite{Aamodt:2011kd}. This work highlights  some of the recent results obtained together with A.~Bialas and K.~Zalewski~\cite{Bialas:2014gca,Bialas:2014boa,Bialas:2015hfa}.

\section{Blast-wave model}
I use the most common version of  the blast-wave model~\cite{Siemens:1978pb,Schnedermann:1993ws,Florkowski:2004tn,Retiere:2003kf}.  At freeze-out,  pions are created at a fixed proper time $\tau=\tau_f=\sqrt{t^2-z^2}$. Then, the  single-particle source function becomes
\ba
w(k,x) = k_0  \cosh\eta e^{-U \cosh\eta + V \cos\phi } f(r),
\label{FT1}
\ea
where $k_0=(m^2+k_\perp^2)^{1/2}$ is the pion energy (with $m$ and  $k_\perp$ being the pion mass and transverse momentum, respectively), while $\eta, \phi$ and $r$ are the space-time rapidity, azimuthal angle and transverse distance from the symmetry axis~\footnote{All irrelevant constants are cancelled in the definition (\ref{FT1})}. I also use the notation $U=\beta k_0\cosh\theta$ and $V=\beta k_\perp\sinh\theta$, with $T=1/\beta$ being the freeze-out temperature. The function $\theta$ describes  the Hubble-like transverse flow $\sinh\theta= \omega r$, with $\omega$ being a parameter~\cite{Chojnacki:2004ec}. 

The function $f(r)$ defines the transverse profile/density of the source. It turns out that the model can describe  the  HBT radii measured by the ALICE collaboration if the function $f(r)$ is taken in the form
\ba
f(r) \sim e^{-(r-R)^2/\delta^2}
\label{shell}
\ea
corresponding to a ''shell'' of the width $\sqrt{2}\, \delta$ and  the radius $R$~\cite{Bialas:2014boa}. Thus, the model contains altogether five free parameters: $T,\; \omega,\;\tau_f,\; R$ and $\delta$, which may depend on the multiplicity of the event. The freeze-out temperature is set equal to $T=$~100 MeV and the width of the shell is taken as $\delta=0.75$~fm. The measured average transverse momentum, taken from CMS~\cite{cmspt},  fixes (for each multiplicity bin) the relation between $\omega R$ and the ratio $R/\delta$.  In this way, one is left with the two free parameters: $R$ and $\tau_f$, which are fitted with the $\chi^2$ method. The model parameters and $\chi^2$ values determined for various multiplicity classes are shown in Table~1. The number of degrees of freedom is~13, $\chi^2$ excludes the first bin in $k_\perp$, while $\chi^2_{\rm tot}$ includes all the bins.
\begin{table}[t]
\begin{tabular}{cccccccc} 
\hline 
mult. class & 1--11 &  12--16  &  17--22 &  23--28 & 29--34 &  35--41 & 42--51  \\ 
\hline \\
$\langle N_c \rangle$        &    6.3  &     13.9  &     19.3  &    25.2  &   31.2  &     37.6  &    45.6  \\
$R$  [fm]          &  1.15  &    1.52   &    1.77   &    1.97  &   2.14  &     2.32  &    2.49  \\
$\tau_f$ [fm]     &  1.90 &   2.18 &  2.37 & 2.50 &  2.63 &  2.74 &  2.80  \\
$\chi^2$    &  0.96&  1.90&  2.89&   4.06&  5.88&   5.45& 11.63 \\
$\chi^2_{\rm tot}$ &18.77&   11.08&   6.34&  4.86&   6.65&  5.79&  11.72
\label{par}
\end{tabular}
\caption{Model parameters determined for different multiplicity classes (the highest multiplicity class has been omitted). }
\end{table}

\section{Correlation functions}
The Bose-Einstein correlation function of two identical particles is defined as the ratio of the two-particle distribution and the product of the one-particle distributions~\cite{LL}. In our model it is  given by the formula~\cite{Bialas:2014gca,Bialas:2014boa}
\ba
C(p_1,p_2) &=& 1+ \frac{\tilde{w }(P_{12}; Q)\tilde{w} (P_{12}; -Q)}{w(p_1)w(p_2)} 
= 1+\frac{|\tilde{w}(P_{12},Q)|^2}{w(p_1)w(p_2)}. \lb{uc}
\ea
Here one uses the definition
\ba
\tilde{w }(P_{12}; Q) &=& \int dx \;e^{iQx}w(P_{12};x), \quad
w(p) = \int dx \;w(p;x), 
\ea
where $P_{12} = (p_1+p_2)/2$ and $Q = (p_1-p_2)$. For boost-invariant and cylindrically symmetric systems,  the average three-momentum of a pair may be set along the $x$ axis (the ``out'' direction). The remaining two orthogonal directions are denoted as the ``long'' and ``side'' directions. The HBT radii are the logarithmic derivatives of the correlation functions calculated along these three directions for $Q \to 0$~\cite{LL}.

\begin{table}[t]
\label{exres}
\begin{tabular}{ccccccc} 
Mult. class & 23--28 & & & & & \\
\hline \\
$P_\perp$ [GeV]
&  $R_{\rm side}$[fm] &  $R_{\rm side}$[fm]
&  $R_{\rm out}$[fm] &  $R_{\rm out}$[fm]  \\ 
&  model &  exp.
&  model &  exp.
 \\ 
\hline \\
0.163 & 1.26 & 1.30$\pm$0.12 & 1.32 & 1.18$\pm$0.17 \\
0.251 & 1.18 & 1.21$\pm$0.10 & 1.12 & 1.15$\pm$0.13 \\
0.349 & 1.09 & 1.06$\pm$0.10 & 0.92 & 0.93$\pm$0.10 \\
0.448 & 1.01 & 0.99$\pm$0.10 & 0.79 & 0.73$\pm$0.10 \\
0.548 & 0.94 & 0.97$\pm$0.10 & 0.70 & 0.63$\pm$0.10 \\
0.648 & 0.89 & 0.91$\pm$0.12 & 0.65 & 0.48$\pm$0.13
\end{tabular}
\caption{An example of our results for the multiplicity class 23--28 for the side and out direction.}
\end{table}

An example of our results, obtained for the multiplicity class 23--28 for the ``side'' and ``out'' directions, is shown in Table~2. The agreement between the model results and the data is very good. Interestingly, one can reproduce the values of the radius $R_{\rm out}$ that becomes smaller than $R_{\rm side}$ for large values of $P_\perp$. This behaviour reminds the HBT puzzle discussed in the context of heavy-ion collisions~\cite{Broniowski:2008vp,Pratt:2008qv}. One can check that in our case, the crucial ingredient of the model, responsible for the correct description of the ratio $R_{\rm out}/R_{\rm out}$, is the ``shell'' structure described by the density profile~(\ref{shell}). This finding is similar to the observation made in \cite{csorgo1} where the data \cite{NA22} has been analysed.
 
\section{Closing remarks}
In this conference contribution I have given further evidence that pions 
produced in high-energy $pp$ collisions originate from a thermally equilibrated,  expanding source. 

Let me note that various attempts to interpret hadron production within thermal and statistical models have been the subject of a very successful collaboration between Krakow theoretical and Warsaw experimental groups, which was initiated by Professor Jan Pluta at the beginning of 2000s. Our intense scientific contacts led to the construction of the Therminator code~\cite{Kisiel:2005hn} and to many papers discussing various aspects of femtoscopy~\cite{Kisiel:2006is}, including the HBT puzzle~\cite{Broniowski:2008vp}. During the WPCF 2015 Workshop we have celebrated Jan's 70th birthday. I hope Jan will remain in good shape, very active, and full of inspiring ideas for our community in the following years. 

{\bf Acknowledgments:} This work has been supported in part by Polish National Science Center Grant No.
DEC-2012/06/A/ST2/00390.

\end{document}